\documentstyle{l-aa}    
\def\skuno{\vskip 20pt}
\begin{document}
\thesaurus{11.03.1; 11.03.4 A2255; 11.09.3; 13.18.1; 13.18.2, 13.25.3}
\title
{The radio and X-ray properties of Abell 2255}
\skuno
\skuno
\author{L. Feretti\inst{1,2} \and H. B\"ohringer\inst{3} \and
G. Giovannini\inst{1,2} \and D. Neumann\inst{3}}

\offprints{L. Feretti}
\institute{
Istituto di Radioastronomia -- CNR, via Gobetti 101, I--40129
Bologna, Italy.
\and Dipartimento di Astronomia dell'Universit\'a, 
Via Zamboni 33, I--40126 Bologna, Italy.
\and Max Planck Institut f\"ur Extraterrestrische Physik, PO Box 1603, 
D--85740 Garching, FRG}

\maketitle

\begin{abstract}

New radio and X-ray data are reported for the rich cluster Abell 2255.
The cluster radio emission is characterized by the presence of a
diffuse halo source, located at the cluster center, and a more peripheral
ridge of radio emission which could be a relic. In addition, 
6 tailed radio sources and a small double are found to be
associated with cluster galaxies. 
At X-ray wavelengths, the cluster shows  elongated emission in the center,
indicating an ongoing cluster merger, with the axis in
the east-west to southeast-northwest direction.
The overall cluster temperature is found from the ROSAT
observation to be $kT = 3.5 (\pm 1.5)$ keV,
while the temperature determined from the Einstein MPC data was
reported to be 7.3 keV. We interpret this difference
as due to the existence in the cluster of several coexisting temperature 
components. 

The merger process can provide energy to maintain the radio halo, but we
stress the need for tailed radio galaxies orbiting at the cluster center
to provide the halo relativistic electrons. 

\end{abstract}

\section{Introduction}

Recent observations of clusters of galaxies have revealed a new and complex
scenario in the structure of the intergalactic medium. 
The clusters are not simple
relaxed structures, but are still forming at the present epoch. Substructures,
commonly observed  in the X-ray distribution of  a high number of rich
clusters (Henry \& Briel 1993, Burns et al. 
1994),  are evidence of hierarchic growth of clusters from the merger
of poorer subclusters.

An important problem in cluster phenomenology regards cluster-wide radio
halos, whose prototype is Coma C (Giovannini et al. 1993, and references
therein). These are extended diffuse radio sources with typical sizes of
0.8-1.2 Mpc (H$_0$=50 km s$^{-1}$ Mpc$^{-1}$) and steep radio spectra. Their
origin and properties are still poorly understood. According to
recent suggestions, the cluster merger process
may play a crucial role in
the formation and energetics of these sources
(see Feretti \& Giovannini 1996 and reference therein). 

We present here radio and X-ray data of the cluster Abell 2255, 
known to contain 
a diffuse radio halo (Jaffe \& Rudnick 1979, Harris et al. 1980,
Burns et al. 1995) and characterized 
by the presence of several tailed radio galaxies. 
It is located at the redshift of 0.0809 (Postman et al. 1992). 
From the optical point of view, this cluster is 
classified as RS type C (Struble \& Rood 1982) and shows a
radial velocity dispersion of 1221 km s$^{-1}$, one of the largest 
observed for any cluster (Zabludoff et al. 1990). Analysis of  the 
spatial and redshift distribution of cluster galaxies (Zabludoff et al. 1990)
shows no evidence of optical subclumps.
At X-ray wavelengths, a temperature of 
7.3 keV is reported by David et al. (1993), and 
no cooling flow has been detected (Edge et al 1992).

In this paper, 
the radio and X-ray observations of A2255 are analysed to derive
the connection between the X-ray and radio properties. We also discuss 
the physical conditions of
the intergalactic medium and of the radio halo and extended radio galaxies.
With the adopted value
H$_0$=50 km s$^{-1}$ Mpc$^{-1}$, 1 arcsec corresponds to 2.01 kpc. 

\section{Radio Data}

\begin{figure}
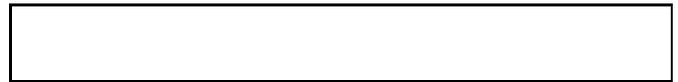
  
\picplace{1.0cm}    
\caption{ Radio map at 90 cm with resolution of 89"$\times$84"(@6$^{\circ}$).
The $\sigma$ noise level in this map is 1.3 mJy/beam.
Contours are --4, 4, 7,10,20, 30, 50, 70 100, 300, 500, 1000 mJy/beam.
Crosses indicate the radio galaxies.}
\end{figure} 

Two full synthesis observations
were obtained with the Westerbork Synthesis Radio Telescope
(WSRT) in January and May 1992, at 90 and 20 cm,
with short spacings of 36 and 72 m, and 36 and 54 m, respectively.
Data were calibrated using the Dwingeloo Westerbork
Astronomical Reduction Facility package  (DWARF) and
further reduced with the  Astronomical Image Processing
System (AIPS), 
following the standard procedure (Fourier inversion,
CLEAN and RESTORE). 

The angular resolutions of the final maps at 90
 and 20 cm are $\sim$85" and $\sim$12", respectively.
In addition to the full resolution map, a  
map degraded to a lower resolution 
was produced  at 20 cm by omitting  the longest 
baselines, in order to enhance the low
brightness structure. A spectral index map was produced comparing
two maps obtained with the same UV coverage.
We also obtained maps of the polarized intensity in the standard way.

\subsection{Diffuse halo} 

The radio image of the cluster center
at 90 cm is presented in Fig. 1. Extended emission is detected from 
cluster galaxies, indicated in 
Fig. 1 by crosses. 
Moreover, the radio emission is 
characterized by the diffuse halo, showing one component  at 
the cluster center (A), and a more peripheral
component of very elongated shape (B). 
This last feature, which 
 has been mapped for the first time by Burns et al (1995), is
 not obviously associated with any cluster galaxy.
 The component A shows here a size much
larger than in the previous maps by Harris et
al. (1980) and Burns et al. (1995), owing to the high sensitivity
of the present map to the
extended regions of low brightness and steep spectrum.
The centroid of the halo region (component A) is 
about RA(J2000)= 17$^h$ 12$^m$ 45$^s$,
DEC(J2000)=  64$^{\circ}$ 04$^{\prime}$ 30$^{\prime\prime}$.

\begin{figure}
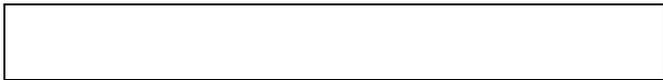
 
\picplace{1.0cm}    
\caption{ Radio map at 20 cm, convolved to the resolution of 70" (HPBW).
The $\sigma$
noise level is 0.1 mJy/beam. Contours are --0.5, 0.3, 0.5, 0.7, 1, 1.5, 2, 3,
5, 7, 10, 30, 50, 100, 150, 200 mJy/beam. Large negative residuals are
due to the lack of short spacings in the UV data.}
\end{figure} 

\begin{table}
\caption{Diffuse source}
\begin{flushleft}
\begin{tabular}{lll}
\hline
\noalign{\smallskip}
  &  Component A & Component B \\
\noalign{\smallskip}
\hline
\noalign{\smallskip}
S - 90 cm (mJy) & 536  & 103  \\
S - 20 cm (mJy) & 43  & 12  \\
Largest Size (kpc)   & 1200 & 965 \\
Luminosity (W) & 1.6$\times$10$^{34}$ & 2.2$\times$10$^{33}$ \\
Volume (kpc$^3$) & 4.7$\times$10$^{8}$ & 3.7$\times$10$^{7}$ \\
u$_{min}$ (erg cm$^{-3}$)  & 2.1$\times$10$^{-14}$ & 1.9$\times$10$^{-14}$ \\
U$_{min}$ (erg) & 2.9$\times$10$^{59}$ & 2.1$\times$10$^{58}$ \\
H$_{eq}$ ($\mu$G) & 0.48 & 0.45 \\
\noalign{\smallskip}
\hline
\noalign{\smallskip}
\end{tabular}
\end{flushleft}
\end{table}

At 20 cm, the diffuse halo is below the sensitivity of the 
full resolution map, but is visible in the  lower resolution map (Fig. 2),
where the signal to noise ratio is improved.
The extended emission in this map is clumpy, and in the northwestern
halo region has the shape of a ring.
This structure is in good agreement
with the VLA map obtained by Burns et al. (1995) at 20 cm and is also
present in our 90 cm map, less clearly because of the larger
beam. 

The flux density, angular size, and intrinsic parameters for the
diffuse source components are given in Table 1.

\begin{figure*} 
\picplace{1.0cm}    
\caption{ Radio maps of the individual radio galaxies, at 20 cm, with
the resolution of about 12.7"$\times$11.6"(@-2$^{\circ}$). 
The noise r.m.s. in the maps is 0.05 mJy/beam.
Contour levels are: -0.3, 0.3, 0.7, 1.5, 3, 5, 10, 25, 50,
100 mJy/beam for 1712+640 and
1713+641;  -0.15, 0.15, 0.3, 0.5, 0.7, 1, 2, 3, 5, 10, 20
for the others.}
\end{figure*} 

\begin{table*}
\caption{Properties of extended radio galaxies}
\begin{flushleft}
\begin{tabular}{lllllll}
\hline
\noalign{\smallskip}
Radiogal. &  WSRT\#  &  RA(J2000) &  DEC & S$_{90}$ & S$_{20}$ & Struct. \\
         &  14W        &  h \ \ m \ \ s 
&  $\circ$ \ \ $\prime$ \ \ $\prime \prime$ & mJy & mJy \\
\noalign{\smallskip}
\hline
\noalign{\smallskip}
B 1711+640 & 113 & 17 \ 12 \ 16.2 & 64 \ 02 \ 06 & 1282.4$^*$ & 10.7 &  Tail \\
B 1712+640  & 115 & 17 \ 12 \ 24.6 & 64 \ 02 \ 08   &    & 303.2 &  Tail \\
B 1712+641  & 118 & 17 \ 13 \ 04.8 & 64 \ 06 \ 59   & 202.4 & 64.8 &  Tail \\
B 1712+638  & 119 & 17 \ 13 \ 16.5 & 63 \ 47 \ 42   & 296.2 & 82.7 &  Tail \\
B 1713+641  & 120 & 17 \ 13 \ 29.3 & 64 \ 02 \ 49  & 597.3 & 276.65 &  Double \\
B 1713+643  & 123 & 17 \ 14 \ 05.6 & 64 \ 16 \ 02   & 28.4  & 7.3  & Tail \\
B 1714+641  & 127 & 17 \ 15 \ 09.0 & 64 \ 02 \ 54   & 186.6 & 57.5 &  Tail \\
\noalign{\smallskip}
\hline
\noalign{\smallskip}
\end{tabular}
\end{flushleft}
$^*$This flux refers to the sum of 1711+640 and 1712+640, which are blended 
in the 90 cm map
\end{table*}

The integrated spectrum of the extended emission between 90 and 20
cm is $\alpha^{90}_{20} \sim$1.7.
The spectral index distribution of the halo between 90 and 20 cm is
difficult to obtain, because the  subtraction of extended radio galaxies
present within the diffuse emission is tricky especially
at 20 cm.
As a general trend, we obtain a fairly constant steep spectrum in the halo
region, with $\alpha>$ 1.5, and no strong
evidence of steepening in the outermost region. An evident steepening is
instead present in the "hole" region, where we derive a spectral index $>$2.
The spectrum of the northern elongated ridge (component B) is
fairly constant and slightly flatter, with spectral index 
around 1.3-1.5.

We made polarization maps of the cluster at both frequencies.
At 20 cm, polarized flux is detected from all the extended radio galaxies, but
not from the diffuse emission. 
This implies an upper limit to the polarized brightness of the halo
of 0.1 mJy/beam, and an upper limit to the polarized 
flux percentage of $\sim$9\%. At 90 cm, no significant polarized flux is 
detected either in the halo or the extended radio galaxies.
This implies an upper limit of $\sim$2\% for the halo polarization percentage.

\subsection {Radio galaxies}

In the full resolution image at 20 cm, five tailed radio sources and 
a small double can be identified
with cluster galaxies, in accord with Harris et al. (1980).
Another tailed radio source, B1713+643, is found to be coincident with
a 18.2$^m$ galaxy of unknown redshift. This source was detected
also by Harris et al. (1980), but was unresolved in their observations.
Owing to its morphology, we favour the hypothesis that this source also is 
a cluster radio galaxy. 

The  cluster radio galaxies showing extended structure
are listed in Table 2, where the position
refers to the optical galaxy. They are 
marked by crosses in Fig. 1, while their high resolution radio images
are presented in Fig. 3.

In the sources 1711+640, 1712+640 and 1712+641 the
nucleus and jets are not resolved. 
The tail transverse size in  1712+640 is rather  constant, indicating
the efficiency of confining effects. In the low
resolution map,
this source shows a sharp bend to the north  (see Fig. 2)  where it merges
into the halo. A similar structure is seen in the tailed radio
 galaxy NGC4869, which
 merges into the Coma cluster radio halo
 and is suggested to be responsible for the relativistic
 electron supply to the halo itself (Giovannini et al. 1993).
 The tail of 1712+641 is very
narrow, with the transverse size unresolved by the 
$\sim$12$^{\prime\prime}$ observing beam.
The source 1712+638 resembles the prototype head-tail galaxy NGC 1265 in
the Perseus cluster (O'Dea \& Owen 1986), but again the transverse
size of the tail is narrow, with a pinch at about 1.5' from the core.
In 1714+641, the twin opposite jets are well visible. They
end in faint hot spots, from which low brightness tails originate. This source
is likely to be affected by projection effects. 

Finally, we remark the peculiarity of  the source 1713+641, which
shows a double structure with separation of $\sim$20$^{\prime\prime}$ (40 kpc).
 At higher resolution (Rudnick \& Owen 1977),  this source shows the
 typical Fanaroff-Riley II structure (Fanaroff \& Riley 1974)  
with a faint core and two symmetric lobes. The
 double radio structure is rare in clusters (Fanti 1984); moreover, this
source is very small for its power of 8 $\times$ 10$^{24}$ W/Hz at 1.4 GHz  
(see the size-power relation of  radio galaxies in
clusters, Feretti \& Giovannini 1993).

\section{X-ray Data}

\subsection {Spatial analysis}

A2255 was observed by us with the ROSAT High Resolution
Imager (HRI) detector, 
in two pointings in January and June 1994, for a total of $\approx$ 32,500 sec. 
A 14,500 sec exposure, obtained with the Position Sensitive Proportional
Counter (PSPC), was taken from the ROSAT archive.
The ROSAT energy  band is 0.1-2.4 keV for both detectors.
The HRI  provides high angular resolution (nominally FWHM $\sim$
2$^{\prime\prime}$),  while the PSPC has an angular
resolution of $\sim$ 25$^{\prime\prime}$ (FWHM) at 1 keV,  
as well as energy resolution. 
A detailed description of the instruments can be found in 
Tr\"umper (1983) and Pfeffermann et al. (1987).
The data analysis has been performed with the EXSAS package (Zimmerman et
al. 1994). Maps of X-ray brightness distribution were produced by binning
the photon events in a two-dimensional grid and then smoothing with a gaussian
filter. The PSPC image was made in the hard band, corresponding to the
energy range 0.5-2.0 keV.

\begin{figure} 
\picplace{1.0cm}    
\caption{ ROSAT/PSPC image of A2255. The image is filtered
with a gaussian of 50$^{\prime\prime}$(FWHM).
The contour levels are 0.025, 0.04, 0.065, 0.1, 0.17, 0.25, 0.4, 0.6 
count s$^{-1}$ pixel$^{-1}$ (1 pixel 
= 15$^{\prime\prime} \times$ 15$^{\prime\prime}$). The 
background is about 0.012 count s$^{-1}$ pixel$^{-1}$.}
\end{figure} 

\begin{figure} 
\picplace{1.0cm}    
\caption{ ROSAT/HRI contours of A2255, superimposed on the
grey-scale image from the digitized Palomar Sky Survey. 
To enhance low brightness structure, 
the X-ray image is smoothed with gaussians of increasing width with 
decreasing count rate, from 10$^{\prime\prime}$ to 50$^{\prime\prime}$(FWHM).
The contour levels are 0.7, 0.8, 0.9,
1, 1.1, 1.2, 1.3, 1.4, 1.5 counts/pixel 
(1 pixel=4$^{\prime\prime} \times$4$^{\prime\prime}$).}
\end{figure} 

Figs. 4 and 5 show PSPC and HRI images of A2255, respectively.
The X-ray emission shows elliptical structure in the PSPC, 
confirmed by the high resolution HRI image.
In the outer low surface brightness region the cluster appears 
to be more symmetric. 
The centroid of the X-ray emission is approximately
RA(J2000)=17$^h$12$^m$45$^s$, DEC=64$^{\circ}$03$^{\prime}$54$^{\prime\prime}$.
The X-ray peak is shifted to the west with respect to this position.

A radial profile of the X-ray surface brightness was 
obtained  by integrating the PSPC counts over concentric annuli 
of $15^{\prime \prime}$, centered on the approximate symmetry 
center. Similarly, a radial profile of the HRI X-ray emission  
was obtained by
integrating the photon counts over concentric annuli of $8^{\prime
\prime}$.
In this averaging process any deviations from symmetry are neglected.
The profiles were fit with a  hydrostatic isothermal model
described by the functional form (Cavaliere \& Fusco-Femiano 1981, 
Sarazin 1986):
$$S(r) =S_0 (1 + r^2/r^2_{\rm c})^{-3 \beta + 0.5}$$
 where S$_0$ is the central surface brightness, $r_{\rm c}$ is the 
core radius, and  $\beta$ is the ratio of the galaxy
to gas temperature. The cluster background is also fit.
The parameters obtained from the fit to the PSPC 
profile are $r_{\rm c}$=4.8$^{\prime} \pm$ 0.4$^{\prime}$ and
$\beta$=0.74$\pm$0.04, 
in excellent agreement with the values $r_{\rm c}$=4.92$^{\prime}$
and $\beta$=0.77 obtained by Jones \& Forman (1984) with the Einstein
Observatory in the 0.5-3 keV energy range.
The profile is shown in Fig. 6. 
The fit to the HRI surface brightness profile is problematic,
because the cluster extends over the whole field of view and therefore
the profile does not reach the background.
However, the  model obtained from the PSPC fit provides a very good
fit also for the HRI data. Therefore, a single $\beta$ model fit
can satisfactorily reproduce the X-ray brightness profile. 

\begin{figure}
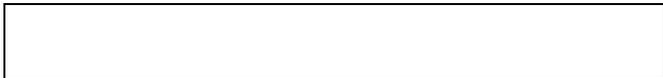
 
\picplace{1.0cm}    
\caption{ Surface brightness profile of A2255 from the PSPC data.
The full line shows the fit of the $\beta$ model. }
\end{figure} 

The asymmetric structure of the cluster is shown in detail if 
the spherically symmetric cluster component corresponding to the best-fitting
azimuthally averaged $\beta$ model is subtracted from the X-ray image.
Apart from dicrete sources,  the main residuum is in the eastern sector.
This decomposition is not perfectly unique, however, since the results 
depend on the choice of the center of symmetry. We have chosen as
symmetry center the X-ray maximum located approximately in the center of the
innermost elliptical contour of the PSPC image, and a bit to the east in
the innermost elongated contour of the HRI image. The fact that the cluster
in the PSPC image appears more elongated to the east and southeast,
with respect to the X-ray peak, now gives rise to the compact structure 
in the residuum shown in Fig. 7.  In the innermost region, there is also
a small region of excess emission lying west of 
the adopted center. At larger radii,
one also observes very low surface brightness emission in the northwest
(opposite the more prominent substructure in the southeast) in which
a cluster of point sources is embedded. Even though this low surface brightness
emission is probably real its significance is not high and it falls below
the first contour in Fig. 7. The overall morphology of the X-ray image
leads to the conclusion that this cluster is another example of merging
subclusters (see also Burns et al. 1995). Compared to the case of A2256 
(Briel et al. 1991) where the infalling subcluster could be well separated,
it is more difficult here to unravel the merger structure. But the
situation suggests that the major merger axis is in the east-west
to southeast-northwest direction.

\begin{figure}
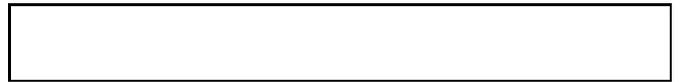
 
\picplace{1.0cm}    
\caption{Residual of the ROSAT/PSPC image, after subtraction
of a symmetric $\beta$ model. No negative contours are plotted, since
they only contain noise and no significant structure. }
\end{figure} 

\subsection {Spectral analysis}

We have analysed the X-ray spectra from the PSPC observation both in the
whole cluster and in several
concentric rings around the cluster center. 
Since the cluster is elongated in the east-west direction, which may also be
the axis of the possible cluster merger, we also 
looked for differences in the eastern compared to the western sectors. 
The X-ray spectra in the
studied areas are background-subtracted using regions in a ring between
25$^{\prime}$ and 33$^{\prime}$ radius. 

The spectra were analysed by fitting 
models of the emission of hot, optically thin plasma
(Raymond \& Smith 1977) with 
the hydrogen column density, $N_H$, 
the plasma temperature, $kT$, the relative
abundance of heavy elements, and the normalization amplitude
as fit parameters. 
The hydrogen column density was found to be generally close to the expected
galactic absorption.
The heavy element abundance of the emitting material was generally fixed to
0.35 times the solar value. Tests with values in the range 0.2 - 0.5 
times the solar
values (the typical range observed for clusters, Ohashi 1995) showed
that the variation of this parameter has little influence on the results for
the other parameters.

The overall temperature found for the cluster is $kT = 3.5 (\pm 1.5)$ keV.
There is no significant difference in the results for the inner region
inside a radius of 5$^{\prime}$ and the outer region to 10$^{\prime}$
or 20$^{\prime}$ radius.
The temperature for this cluster determined from the MPC data
obtained with the EINSTEIN observatory (David et al. 1993)
gave a value of $7.3 {+1.7 \brack -1.1}$
keV, which is consistent with the high velocity dispersion of this cluster. 
There seems to be a clear discrepancy between these results, unless 
the cluster has several coexisting temperature 
components. Analysis made with data of
the EINSTEIN MPC, which has a higher energy 
window, is more sensitive to the cut-off of the X-ray continuum spectrum
towards high energy. Therefore the highest temperature component is most 
sensitively detected. The ROSAT energy window (0.1 - 2.4 keV) does not
cover the energy range of the spectral cut-off, and the analysis of the 
PSPC data tends to give more weight to the low temperature components.
Unfortunately, the limited energy 
discrimination of the ROSAT PSPC and the limited photon statistics of the
present data set does not allow for a meaningful multitemperature fit,
which would greatly increase the number of free fitting parameters. Therefore,
we can only tentatively conclude that the discrepancy between the two 
instruments may imply a multiphase cluster medium, with
unresolved temperature variations.
This expectation would also be in line with the interpretation of the cluster
as being in a state of merging.

Burns et al. (1995) found an even lower temperature of 
$1.9 {+2.4 \brack -0.4}$ in the central 5$^{\prime}$ radius region, and
interpreted this as a temperature drop to the center. In the light
of the above results, this is not due to a spatial temperature variation
and is consistent with our value within the errors.

 A comparison of spectra
of the two sectors in different ring areas shows that the spectra from the
eastern region are significantly harder than those from the western parts.
It is very difficult, however, to assign a temperature difference to this
difference in hardness, because of 
variation also in the absorbing hydrogen column density  and 
in the heavy element abundance.       

\section{Discussion}

\subsection{Halo origin}

The diffuse halo in A2255 is comparable in size to the prototype halo
source in the Coma cluster, Coma C (Giovannini et al. 1993), but
is more irregular in the brightness distribution. A map of the diffuse
radio emission at 90 cm,
after subtraction of the imbedded extended sources, is shown 
superimposed on the X-ray image in Fig. 8. It is evident
that the main radio  halo is located at the center of the X-ray 
brightness distribution, as can also be deduced from the positions of the 
halo centroid (sect. 2.1) and the X-ray centroid (sect 3.1).

\begin{figure}
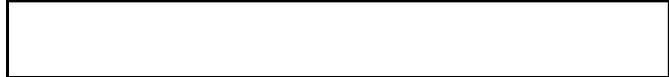
 
\picplace{1.0cm}    
\caption{ Contours of the halo   image at 90 cm, after subtraction of discrete 
sources, 
superimposed on the greyscale X-ray PSPC image.
Contour levels are --4, 4, 7, 10, 15, 20, 25, 30 mJy/beam.}
\end{figure} 

The northern ridge (component B) is at the boundary of the X-ray
emission, and is not obviously related to any X-ray substructure. It is
also unclear if it is related to component A.  It could be located
in a cluster peripheral region, at a true distance much larger than
the projected distance, and could be classified as a relic source.

The properties of radio halo and relics in clusters have been
recently reviewed by Feretti \& Giovannini (1996). While the origin of relics
is still puzzling, the existence of central halos
is likely to be associated with a cluster merger process,
providing  the energy necessary for 
particle reacceleration and magnetic field amplification.
The merger hypothesis alone, however, cannot explain the statistics 
of radio halos;
radio halos are rare, while mergers in clusters seem common. In the case of
Coma-C, Giovannini et al. (1993) suggested that the  tailed radio galaxy
NGC4869, orbiting around the cluster center, is responsible for the
relativistic electron supply. The requirement of 
tailed radio galaxies residing at the
cluster centers as the origin of relativistic particles could explain the
rarity of halo type radio sources. 

A2255 fits the previous picture, owing to the existence of a
cluster merger and the presence of three tailed radio galaxies at the cluster
center, within the halo region. The halo is not spherically symmetric, but
elongated in the east-west direction, i.e., in the merger direction. This
structure is therefore a strong point in favour of a connection between the
merger and the radio halo. Moreover, the halo spectrum shows an evident
steepening in the northwestern region (see Sect. 2.1), i.e., opposite to the
merger. Here, the electron reacceleration is likely to be less efficient. We
note here that the double radio galaxy 1713+641 lies exactly in the merger
region. An attractive possibility to explain its small size and lack of 
distortion is that this is a young source triggered by the merger. 

The orientations of tails of the tailed radio galaxies 
with respect to the cluster center provide clues to 
the distribution of galaxy orbits within
the cluster (O'Dea et al. 1987). The three tailed radio
galaxies embedded within the halo radio emission, i.e., 1711+640,
1712+640, and 1712+641, actually show tails
oriented at random angles with respect to the cluster center, and
are therefore likely to be in random orbits at the cluster center. They
could be responsible for the deposit of relativistic particles radiating in 
the halo, as discussed above.

\subsection {Cluster Mass}

Under the simple assumptions of hydrostatic equilibrium in the intracluster 
medium and spherical symmetry of the cluster it is possible to determine the
total cluster mass from the gas density and temperature distribution. 
The validity
of these  assumptions was studied in N-body/hydrodynamic simulations of cluster
formation by Schindler (1996) and Evrard et al. (1996). These authors
in particular studied how well the cluster mass determined under
these assumptions is consistent with the true mass of the model cluster
for cases where the cluster is approaching the equilibrium configuration.
In A2255 the
merging subcluster probably has already fallen far into the center of the
main cluster and passed the central region of
the main cluster. This intepretation is implied by the fact that
the two bright galaxies are no longer coincident with a maximum
of the X-ray surface brightness. Burns et al. (1995) 
also come to this conclusion
after a comparison of the X-ray and optical appearance of the cluster
with a set of possible model clusters from N-body/hydrodynamical
simulations. In this configuration, the cluster mass determined
under the assumption of hydrostatic equilibrium agrees within about
20\% with the true cluster mass (except for the radial
region around the location of the major outgoing thermalizing
shock wave, Schindler 1996).
Therefore we can get a good account of the total cluster mass by using
the above obtained information on the density distribution
and the temperature of the intracluster plasma. The errors due to
deviations from symmetry and hydrostatic equilibrium are in this case
very likely smaller than the uncertainties introduced due to the
imprecise knowledge of the temperature profile.

\begin{figure}
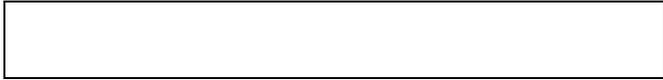
 
\picplace{1.0cm}    
\caption{Distribution of the cluster mass.
The dashed and dotted curves represent respectively the total gas mass of the 
X-ray emitting gas, and the gravitating mass, obtained by
assuming hydrostatic equilibrium. Full curves show the uncertainty
range for the gravitating mass. }
\end{figure} 

To determine the cluster mass, we use the gas density profile
determined from the $\beta $ model fit (shown in Fig. 6) and
gas temperatures in the range of 3.5 to 7 keV. For the temperature
profile, we assume a set of models including isothermal models and
polytropic models with an index in the range 0.9 to 1.3, and the
temperatures fixed to the measured range (3.5 to 7 keV) at the
core radius. The resultant uncertainty range for the gravitating mass
profiles are shown in Fig. 9, with an isothermal model for
$T_{gas} = 5$ keV. The gas mass profile is also shown. At
1 $h_{50}^{-1}$ Mpc the total mass of the cluster is
in the range $1.6 - 3.7 \cdot 10^{14}$ M$_{\odot}$
(consistent with the results by Burns et al. 1995), and the gas mass
fraction is in the range 19 - 44\%. 
If the results are extrapolated to the radius of 3 $h_{50}^{-1}$
Mpc, within which a cluster of the mass of A2255 is expected to be
virialized   (e.g., Gunn \& Gott, 1972),  one finds values of
$M_{grav} = 0.45 - 1.3 \cdot 10^{15}$  M$_{\odot}$, and for the  gas
mass fraction, a range of 33 - 100 \%. Typically, the gas   mass fraction
in clusters is of 10 - 30\%   (e.g. B\"ohringer 1995). Therefore, it is
likely that in A2255  the true mass of the cluster is closer to the upper
limit of the present calculation.  This is consistent with the existence in
the intracluster plasma  of a bulk component with temperature near 7 keV,
in addition to the low temperature component detected by ROSAT. 

\subsection{Radio source confinement}

\begin{figure}
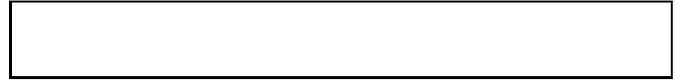
 
\picplace{1.0cm}    
\caption{Allowed range of the pressure of the X-ray 
emitting gas versus the radial 
distance from the cluster center (dashed region), for gas temperatures
between 3.5 and 7 keV. The 
equipartition pressures in the tailed
radio galaxies (dots) and in the radio halo and relic (dashed lines)
are also given.}
\end{figure} 

The gas density and temperature obtained from the X-ray data
allow us to derive the thermal pressure of the intergalactic medium. 
This can be compared with  the equipartition pressures within the 
radio emitting regions, to obtain information about the confinement of 
the radio sources. The three outermost tailed radio galaxies
(1712+638, 1713+643, and 1714+641) are located outside the contours
of the X-ray emission and  therefore in a region of very low gas density. 
From the profile of Fig. 6,
however, it is evident that all these sources 
are still in a region where the X-ray brightness is above the background.
With the parameters obtained from the fit to the observed
brightness profile, a central 
electron density n$_0$ = 1.76 10$^{-3}$ cm$^{-3}$ was obtained.
Pressure profiles were derived for temperatures of 3.5 and
7 keV.

In the computation of equipartition parameters for the radio emitting
regions, we used standard formulae (Pacholczyk 1970) with the
standard assumptions of a low frequency cutoff of 10 MHz, a high frequency
cutoff of 100 GHz, equal energy density in protons and electrons, and a
filling factor of 1. The values of equipartition magnetic field and minimum
internal pressures were computed for the radio halo, for the relic, 
and for the outermost regions of the tailed radio galaxies.

The comparison between the pressure of the X-ray gas and 
the equipartition pressures in the radio emitting regions
is given  in Fig. 10.
As already found in other cases (Feretti et al. 1992, 1995,
R\"ottgering et al. 1994),
the equipartition pressures of the radio sources are systematically 
lower by a factor of $\sim$10 than the 
corresponding thermal pressures of the ambient gas. 
This result implies either that the numerous assumptions used in the
calculation of the equipartition pressure are not valid, or that there is
a real deviation from the equipartition conditions. 

The imbalance between pressures in  the diffuse halo and the ambient 
gas is more dramatic, being a ratio of about 1000 at the cluster center. 
To explain this, we cannot simply relax the assumptions made
for the computation of equipartition energy, but we may invoke another
pressure component, such as thermal gas mixed with the relativistic
particles. This is confirmed by the presence of 
both radio and X-ray emission from the same halo region.

\section{Conclusions}

Abell 2255 is characterized in the radio domain by
several galaxies, and by the presence of diffuse emission
consisting of a halo at the cluster center and a more peripheral 
elongated ridge which we consider a relic.
 At X-ray wavelenghts, the cluster shows an asymmetric structure, 
suggestive of a merger with the axis in
the east-west to southeast-northwest direction.

 The existence of a merger is confirmed by the
spectral analysis, which enhances a cluster component of temperature
$kT = 3.5 (\pm 1.5)$ keV, in addition to the higher temperature
component of 7.3 keV detected with the Einstein MPC data. A
multiphase intracluster medium seems therefore to be associated with
this cluster. 

The merger process can provide energy to maintain the radio halo. 
We also stress the need for tailed radio galaxies orbiting at the cluster center
to provide the halo relativistic electrons. In the cluster A2255,
there are at least three  tailed radio galaxies at the cluster
center, which are likely to be orbiting at the cluster center.
They could be
responsible for the continuous supply of relativistic particles to the halo.

 The X-ray brightness profile, obtained by averaging azimuthally the
count rate in concentric rings around the symmetry center, was fit
with the $\beta$ model, to derive the central density and pressure
of the X-ray gas. 
The internal equipartition pressure in the extended 
lobes of the cluster radio galaxies is 
lower by an order of magnitude than the 
corresponding pressure of the ambient gas.
The internal pressure in the halo is lower than the external one
by about 3 orders of magnitude.

\begin{acknowledgements}
We thank Ger De Bruyn for his help in the radio data calibration, the ROSAT
team for providing the processed data, and the EXSAS team for the tools
for the data reduction. Thanks are due to Jon Aymon for his careful
reading of the manuscript.
L.F. acknowledges the MPE of Garching for
hospitality and partial financial support.

The WSRT is operated by the Netherlands Foundation for Radio
Astronomy with the financial support of ZWO.
\end{acknowledgements}


\begin{thebibliography}{}


\bibitem{} B\"ohringer, H.,  1995, Seventeenth Texas Symp., 
Annals of the New York Academy of Sciences,
  Eds. H. B\"ohringer, G.E. Morfill, J.E. Tr\"umper, Vol. 759, p. 67

\bibitem{} Briel, U.G., Henry, 
J.P., Schwarz, R.A., et al., 1991, A\&A, 246, L10

\bibitem{} Burns, J.O., Rhee, G., Owen, F.N., Pinkey, J., 1994, ApJ 423, 94

\bibitem{} Burns, J.O., Roettiger, K., Pinkney, J., Perley, R.A., 
  Owen, F.N., Voges, W., 1995, ApJ 446, 583

\bibitem{} Cavaliere, A., Fusco-Femiano, R., 1981, A\&A, 100, 194

\bibitem{} David, L.P., Sly, A., Jones, C., Forman, W., Vrtilek, 
   S.D., Arnaud, K.A., 1993, ApJ 412, 479

\bibitem{} Edge, A.C., Stewart, G.C., Fabian, A.C.,  1992, MNRAS 258, 177

\bibitem{} Evrard, A., Metzler, C., \& Navarro, J., 1996, (preprint)

\bibitem{} Fanaroff, B.L., Riley, J.M., 1974, MNRAS 167, 31P

\bibitem{} Fanti, R., 1984, In:  Clusters and Groups of Galaxies, Eds.
Mardirossian F., Giuricin, G., 
Mezzetti M., Reidel P.C., Dordrecht, p.185

\bibitem{} Feretti, L., Perola, G.C., Fanti, R., 1992, A\&A 265, 9

\bibitem{} Feretti, L., Giovannini, G., 1993, A\&A 281, 375

\bibitem{} Feretti, L., Fanti, R., Parma, P., Massaglia, S., Trussoni, E.,
  Brinkmann, W.,  1995, A\&A 298, 699

\bibitem{} Feretti, L., Giovannini, G., 1996, In: Extragalactic Radio
Sources, IAU Symp. 175,  Eds. R. Ekers, C. Fanti \& L. Padrielli, Kluwer Academic
  Publisher, in press

\bibitem{} Giovannini, G., Feretti, L., Venturi, 
  T., Kim, K.-T., Kronberg, P.P., 1993, ApJ 406, 399

\bibitem{} Gunn, J.E., Gott, J.R.III, 1972, ApJ 176, 1

\bibitem{} Harris, D.E., Kapahi, V.K., Ekers, R.D., 1980, A\&AS 39, 215

\bibitem{} Henry, J.P., Briel, U.G., 1993, Adv. Space Res. 13, (12)191

\bibitem{} Jaffe, W.J., Rudnick, L.,  1979, ApJ 233, 453

\bibitem{} Jones, C., Forman, W., 1984, ApJ 276, 38

\bibitem{} O'Dea, C., Owen, F.N., 1986, ApJ 301, 841

\bibitem{} O'Dea, C., Sarazin, C.L., Owen, F. N., 1987, ApJ 316, 113

\bibitem{} Ohashi, T., 1995,  Seventeenth Texas Symp., 
Annals of the New York Academy of Sciences,
  Eds. H. B\"ohringer, G.E. Morfill, J.E. Tr\"umper, Vol. 759, p. 217

\bibitem{} Pacholczyk, A.G., 1970, Radio Astrophysics, Freeman and 
  Co., San Francisco.

\bibitem{} Pfeffermann, E., Briel, U.G., Hippmann, H., et al., 1987,
Proc. Society for Photo-Optical Instrumentation Engineers (SPIE) 733, 519

\bibitem{} Postman, M., Huchra, J.P., Geller, M.J., 1992, ApJ 384, 404

\bibitem{} R\"ottgering, H., Snellen, I., Miley, G., et al., 1994, 
  ApJ,  436, 654

\bibitem{} Raymond, J.C., Smith, B.W., 1977, ApJS 35, 419

\bibitem{} Rudnick, L., Owen, F.N., 1977, AJ 82, 1

\bibitem{} Sarazin, C.L., 1986, Rev. Mod. Phys. 58, 1

\bibitem{} Schindler, 1996,  A\&A, 305, 756

\bibitem{} Struble, M.F., Rood, H.J., 1982, AJ 87, 7

\bibitem{} Tr\"umper, J., 1983, Adv. Space Res. 2, 241

\bibitem{} Zabludoff, A.I., Huchra, J.P., Geller, M.J., 1990, ApJS 74, 1

\bibitem{} Zimmermann, H.U., Becker, W., Belloni, T.,
  D\"obereiner, S., Izzo, C., Kahabka, P., Schwentker, O., 1994,
  EXSAS User's Guide, MPE Report 244

\end{thebibliography}
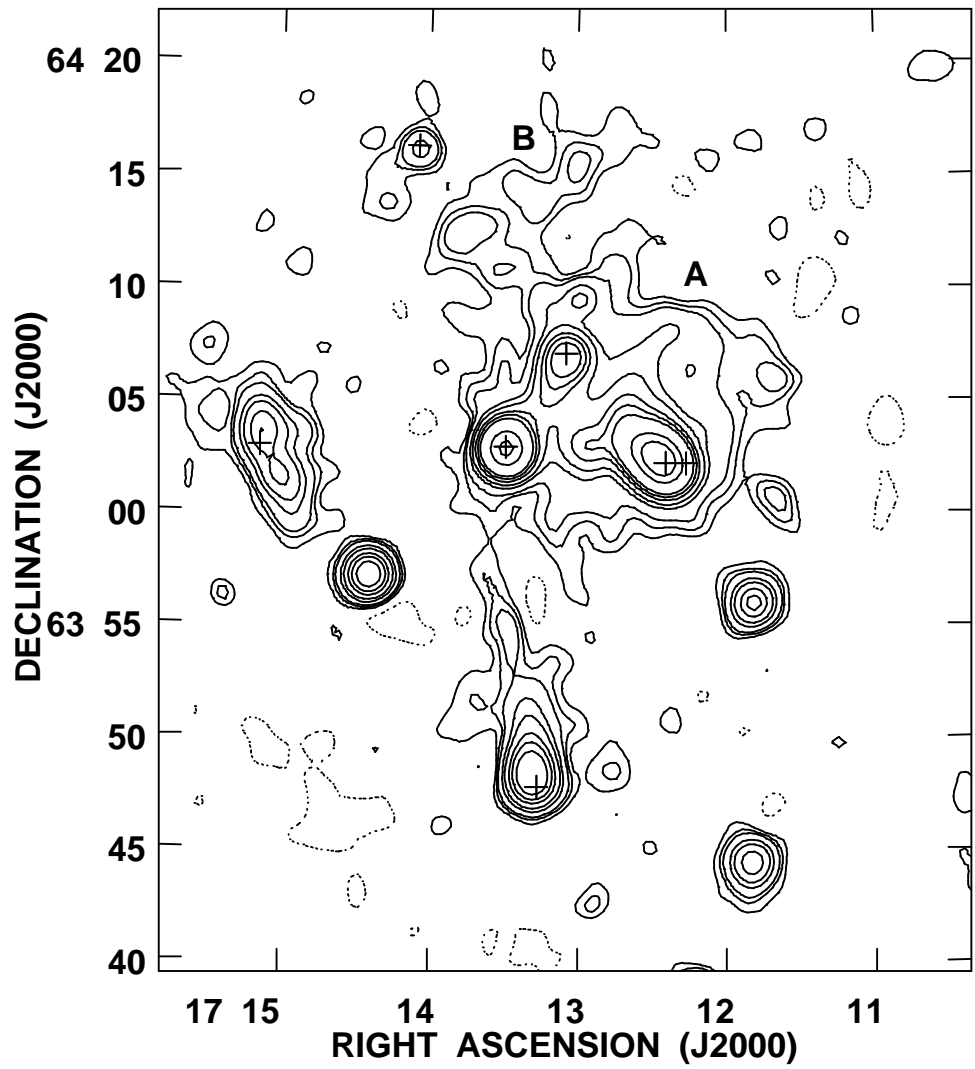
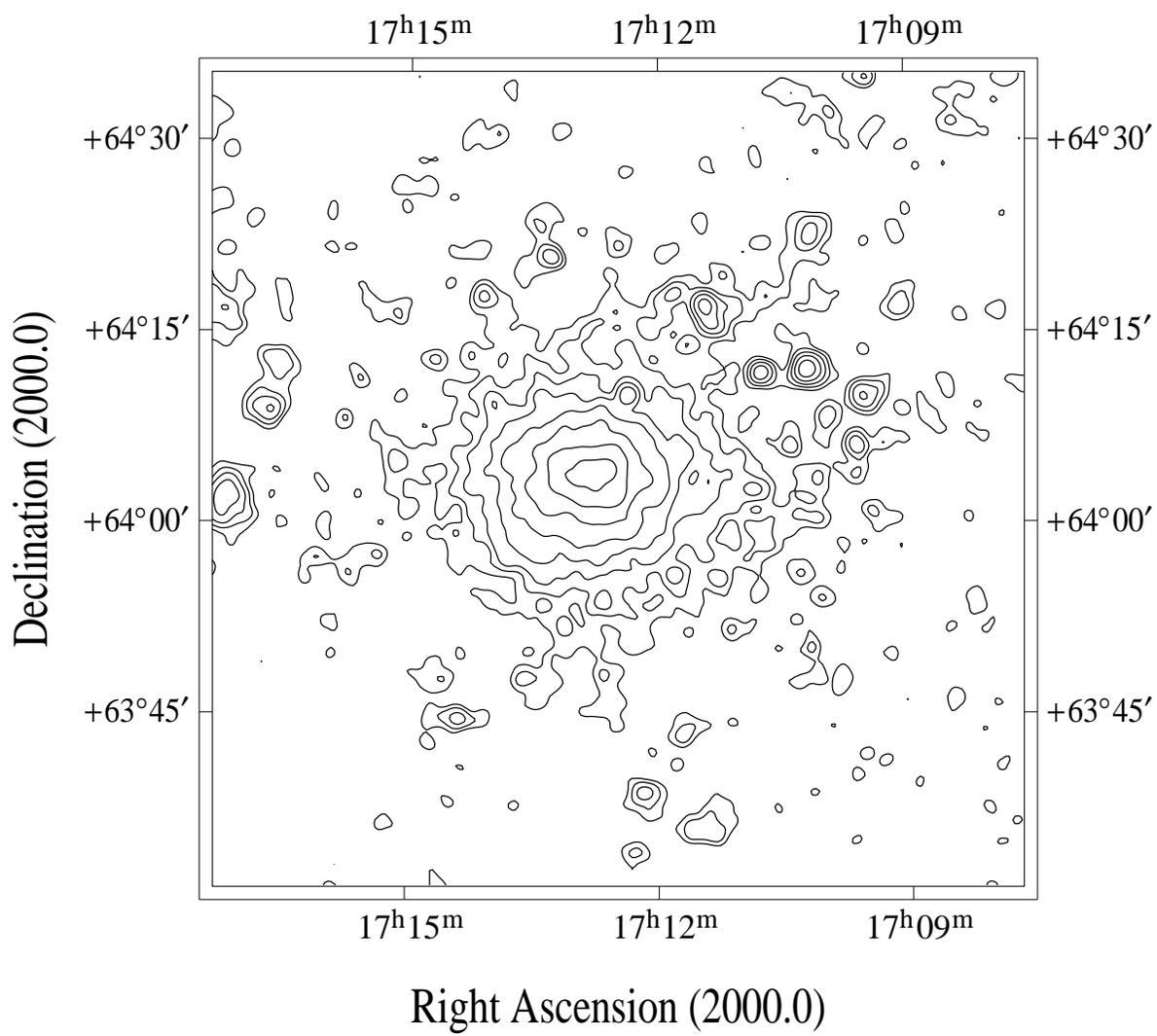
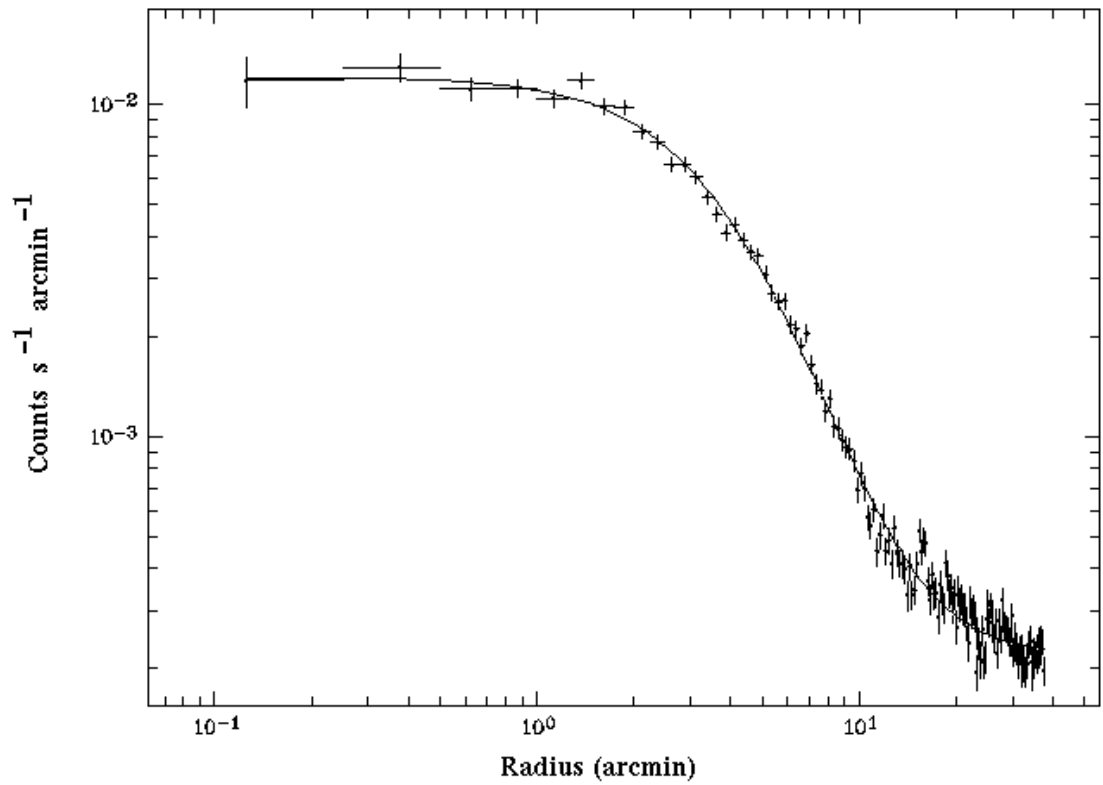
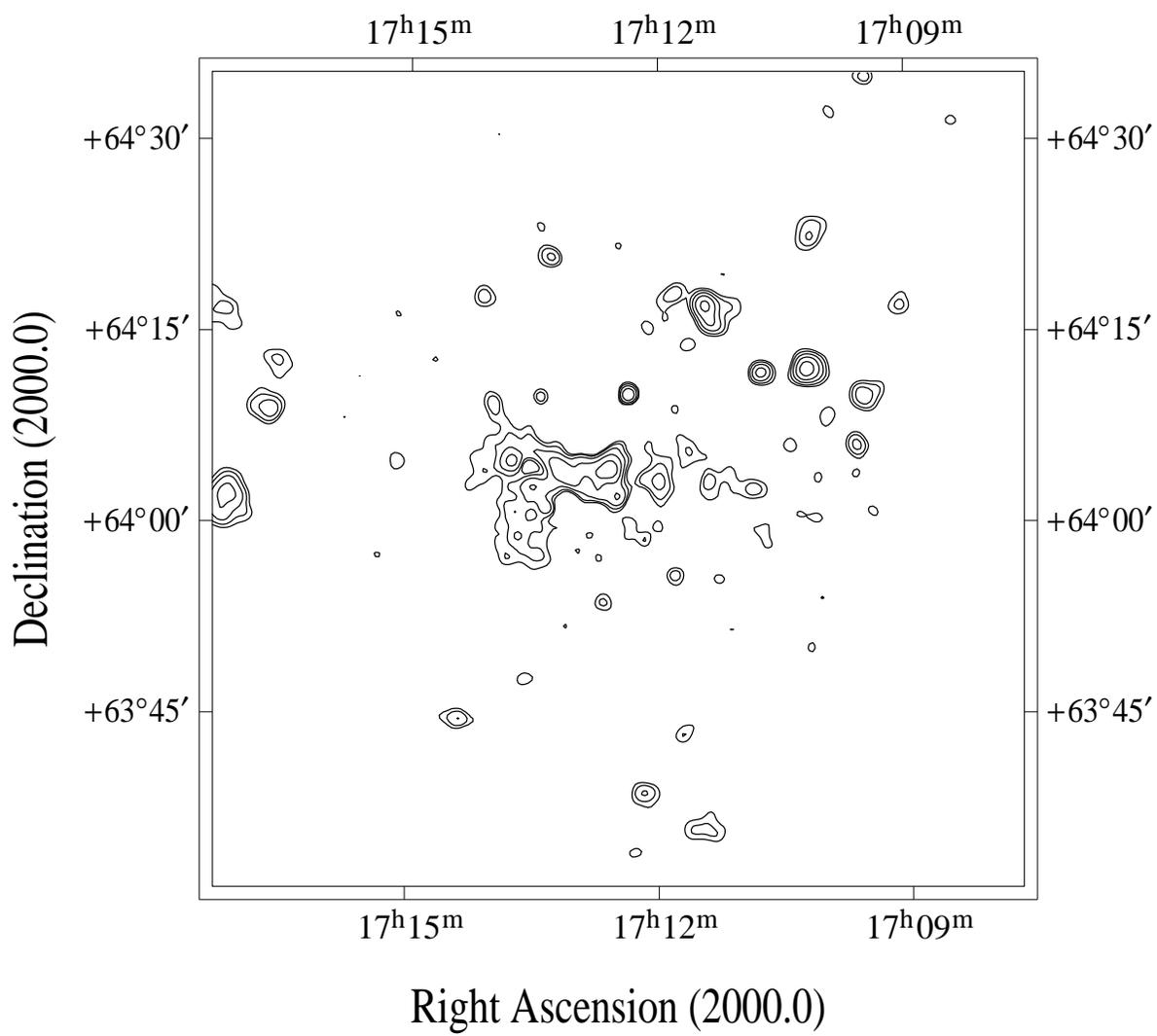
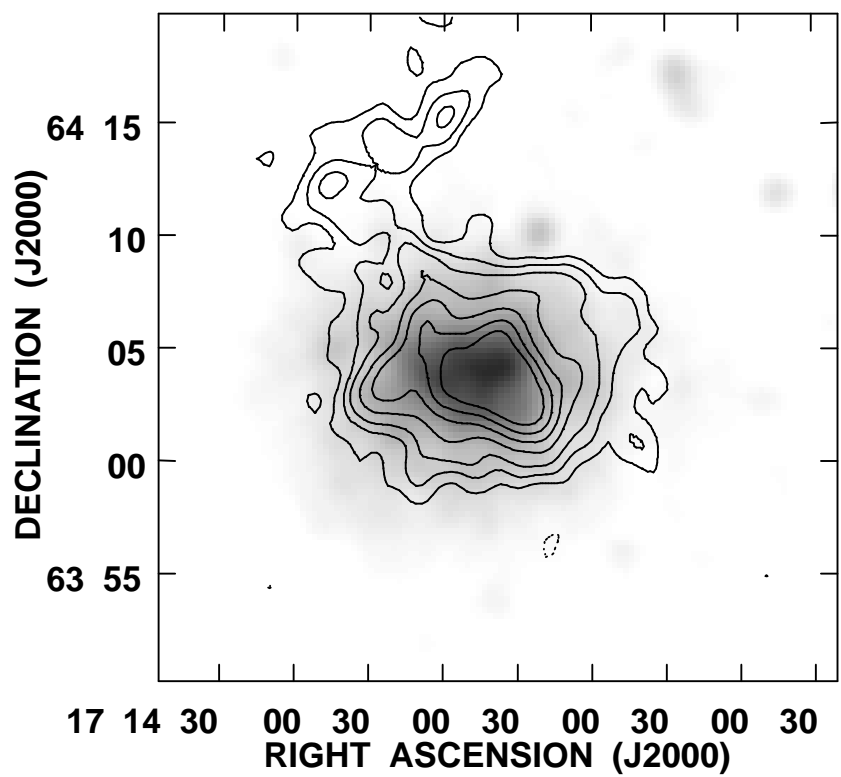
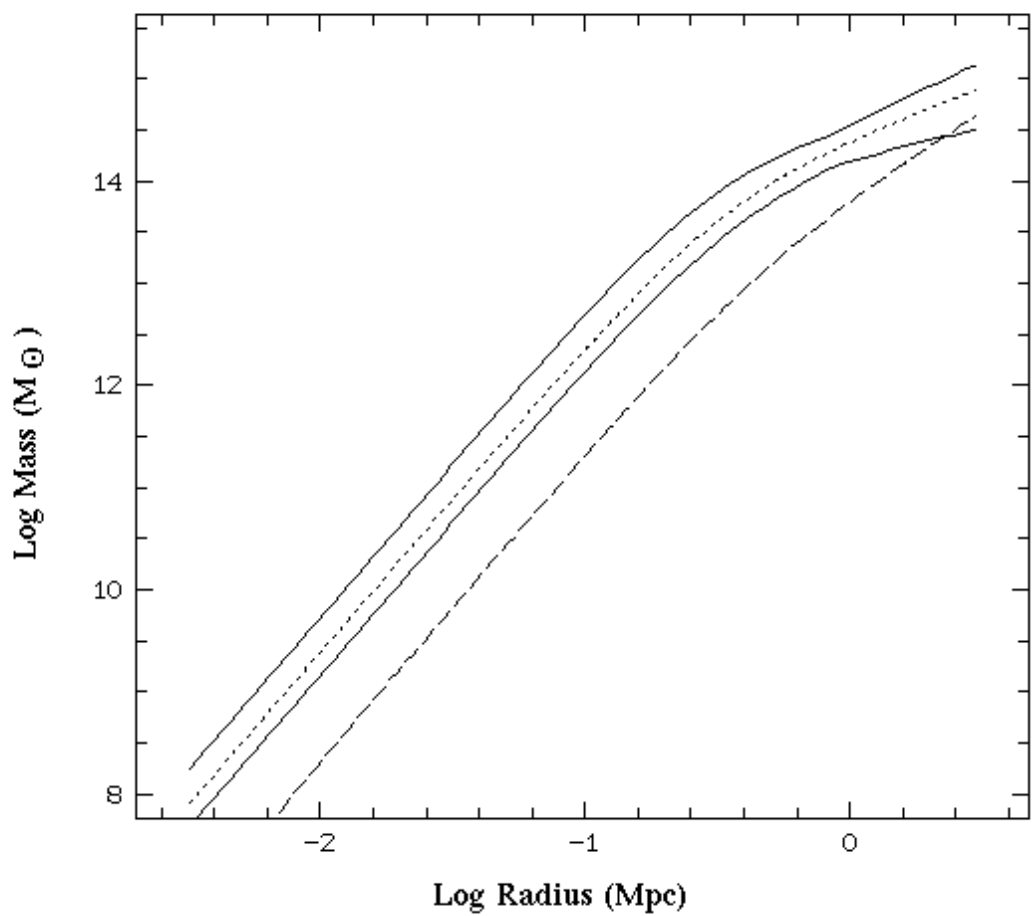
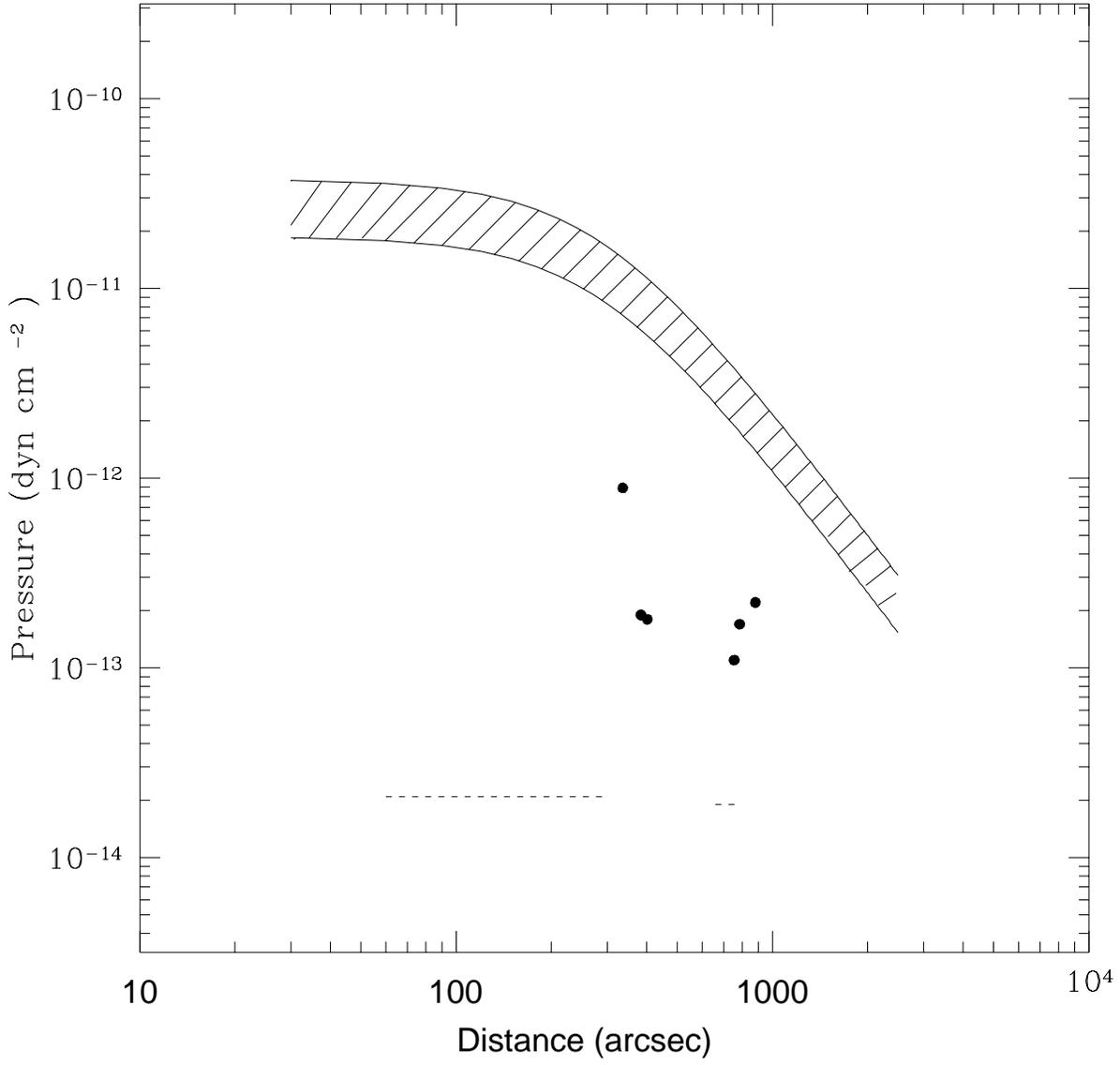
\end{document}